# Search for potential precursors for Si-atomic layer deposition- a quantum chemical study

P. Vajeeston,[a*] H. Fjellvåg,[a] and O. Nilsen,[a]

[a]*Department of Chemistry, Center for Materials Sciences and Nanotechnology, University of Oslo, P.O. Box 1033 Blindern, N-0315 Oslo, Norway*

**Abstract**

Thin film of silicon is an interesting material for many technological applications in electronic industry and in energy harvesting technologies, but requires a method for controlled growth of thin films. The purpose of this study is to screen a wide variety of Si content precursors for Si ALD reactions using state-of-the-art density-functional calculations. Among the studied 85 Si content precursors we found that $C_7H_{12}OSi$ –Methoxy-trivinyl-silane and $C_7H_9NSi$ –Benzyliminosilane show positive indications for ALD reactivity for Si deposition. We believe that this finding will be helpful to develop low-cost, high-energy efficiency thin-film solar cells for future scale up implementation in photovoltaics.

PACS: 82.30.Nr - formation in chemical reactions,

PACS: 46.55.+d Abrasion mechanics,

31.15.AE

*Corresponding author. E-mail address: ponniahv@kjemi.uio.no; Phone: +47 22855613

http://www.folk.uio.no/ponniahv

## Introduction

One of the key technologies with the great scope of reducing greenhouse gas emissions is the solar cell, photovoltaics (PV). Critical factors for bringing PV even stronger into the market are the reduced cost and the increased energy efficiency. The current PV market is based on 300 μm thick silicon wafers cut from large ingots. The current trend is to further reduce the thickness, and hence the cost, of the material. However, due to the finite size of the cutting tools, the proportion of material wasted in the cutting process is increasing to substantial amounts (~ 50%). Alternative methods to produce thin layers of silicon are therefore sought. One concept is to reduce the silicon thickness down to 10-20 μm, which would also enable new designs such as flexible solar cells. Furthermore, the deposited Si would enter as a nucleation or capping layer enabling deposition on inexpensive supports (like steel, glass, etc) in production for PV devices. The development of low cost, high efficiency processes for deposition of Si in solar cells is a key to future scale up of PV.[1] Materials of such 10-20 μm thickness cannot be made with the current top-down approach but require the deposition of the materials on a support. Silicon can be deposited in numerous alternative manners such as physical and metal-organic chemical vapor deposition processes (PVD and MOCVD, respectively) providing suitable growth rates. Nanoparticles of silicon can also be deposited through wet chemical methods forming a basis for radically new design of solar cells. However, in order to achieve the required properties, such as adherence and conformal coating, it is necessary to use a support which is reactive towards silicon.

Atomic layer deposition (ALD) has emerged as an important technique for depositing thin films for a variety of applications. Miniaturization in the semiconductor industry has led to the requirement for atomic level control of the thin film deposition. Miniaturization has produced very high aspect structures that need to be conformally coated. Any other thin film technique can approach the conformality achieved by ALD on high aspect structures. The ALD of single-element semiconductors such as Si and Ge can also be deposited using hydrogen radical-enhanced ALD. Studies of Si ALD using $SiH_2Cl_2$ and H radicals have demonstrated the self-limiting nature of Si ALD growth versus both $SiH_2Cl_2$ and hydrogen radical exposures.[2, 3] The surface chemistry for Si ALD is based on the desorption kinetics for $H_2$, HCl, and $SiCl_2$ from silicon surfaces. $H_2$ desorbs at 535 °C,[4, 5] HCl desorbs at 575 °C,[4] and $SiCl_2$ desorbs at 725 °C[4, 6] from silicon surfaces. A Si ALD growth per cycle of ~1.6 Å was also observed between 550 and 610 °C. But, this temperature is too high for the fabrication of new devises/materials and novel structures such as Si-Ge hetero-junctions or super-lattices, because of the inter-diffusion of Si and Ge. At lower temperatures, the Si ALD growth per cycle decreases as a result of incomplete surface reactions. Si is very reactive and easily reacts with oxygen, forming silicides from oxide substrates (like $SiO_2$) or metallic substrates. Consequently, the nucleation of Si ALD is very difficult. The nucleation problems have limited the surface chemistry for Si ALD. The purpose of this study is to find a potential candidate for Si ALD using state-of-the-art density-functional calculations.

**Computational details**

All calculations have been performed using the atomic orbital density-functional theory (DFT) method as implemented in the $DMol^3$ (MaterialStudio5.5)[7, 8] Double numerical polarized (DNP) basis set that includes all occupied atomic orbitals plus a second set of valence atomic orbitals and polarized *d*-valence orbitals was employed. For exchange and correlation we applied the gradient corrected approach using the generalized gradient approximation (GGA) functional following the approach suggested by Perdew-Burke-Ernzerhof (PBE).[9, 10] It was shown by Delley[11] that the PBE functional with the efficient DNP numerical basis set gives enthalpies of formation for a large set of tested compounds and molecules from the NIST database closer to the experimental values. The estimated error was found to be lower than that obtained with the hybrid B3LYP/6-31G** functional.[11] Self-consistent-field convergence criterion was set to the root-mean-square change in the electronic density to be less than $1\times10^{-6}$ electron/Å$^3$. The convergence criteria applied during geometry optimization were $2.72\times10^{-4}$ eV for energy, 0.054 eV/Å for force and 0.005 Å for displacement. For all the optimized molecular structures we performed frequency analysis to check whether the obtained structure was a true minimum and only the ground state structures are analysed in this paper. In general, quantum simulations of surface reactions, such as those of ALD on the growing surface, can be performed using either a finite cluster or periodically extended supercell model of the reacting surface site. Here, for each Si system, we have used in first place a finite cluster approach. If the reactivity computed on the cluster was found to be positive, the extended model (periodic) was applied to simulate the Si ALD surface reactions.

For the surface calculations, a (100)-slab has been cut from the bulk crystal (FCC Si) structure, previously optimized with respect to stress and strain. On the so-obtained slabs, all atoms have been allowed to relax minimizing the forces acting on them. A thick vacuum region was included to prevent slab-to-slab interaction along the normal direction to the surface. We found that a thickness of 12 Å was sufficient to achieve the energy convergence below the 1 meV/atom threshold. In order to obtain the asymmetric dimer reconstruction at the Si (100) surface we have used large atomic displacement factor for the surface atoms. Once the reconstructed Si (100) was obtained the OH groups are added to the surface dimers and the system was fully relaxed. For consistency of the computed values we have used similar computational parameters for the bulk and the cluster calculations in the DMol$^3$ program. It should be noted that all the formation energies reported in this manuscript (both in the cluster model and in the periodic surface model) are at zero Kelvin. In order to compare the reaction energy between the cluster model and the surface 2D model at 300K, the phonon contribution to the total energy for the 2D model is needed. This is a challenging task due to the huge computational effort involved. However we observed that the thermal contribution (zero-point energy) to the reaction energy in the cluster model calculation is less than 15 kJ/mol (at T= 300K). Given that the energies per formula units are very close in the cluster and 2D model we expect the phonon correction to be negligible with respect to the zero-point energy. We therefore speculate that the picture of the relative reactivity for the analyzed precursors should not change significantly when the effect of the temperature is taken into account.

**Results and discussion**

In order to identify the potential precursor for Si ALD reactions we have considered 85 Si content precursors for our initial screening process (see Table I). Since there is no unique criteria for the selection of the precursor molecule for the screening, the aforementioned molecules are chosen randomly among the precursors in the Chemspider database[12] with the lowest molecular formula (less than 40 atoms in the precursor). The precursors correspond to the lowest energy structure of the given molecular formula. The chosen precursor list includes, besides Si(IV)-based compounds, also some of the smallest Si(II)-precursors[13, 14] and Si(III)-precursors.[15]

The interaction of the aforementioned (see Table I) precursor molecules with the Si(100) H-terminated (open bonds on the surface are terminated by H), surface and the Si(100) OH-terminated (open bonds on the surface are terminated by OH group; hydroxylated) surface is also modeled. In order to speed up our screening procedure we have used a simplified $Si_9H_{12}$ cluster model (Fig1) that corresponds to Si(100)-2×1 surface. This one-dimer cluster is frequently used to explore the reaction mechanism on the hydroxylated/H-terminated Si(100)-2×1 surfaces.[16] The cluster approach is based on the predominantly localized bonding of the Si(100)-2×1 surface.[16] In order to mimic the three layers of Si ALD process we have considered the Si(100)-2×1 three dimer surface (see Fig.

1). The $Si_{21}H_{20}$ three-dimer cluster consists of four layer silicon atoms where the top six silicon atoms form the surface dimers. The remaining fifteen silicon atoms form three subsurface layers that are terminated by hydrogen atoms to prevent unrealistic charge transfer. The two hydroxyls/H-terminated moieties of $Si_{21}H_{20}$–$(OH)_6$/$Si_{21}H_{20}$–$(H)_6$ will be the active surface sites for the reaction with the incoming Si-based precursor. The aforementioned precursors and their reaction mechanism on the hydroxylated/H-terminated Si(100)-2×1 three-dimer surface are studied (Fig.1). If the chosen precursor have a positive indication for the Si deposition (i.e the formation energy, defined as the energy difference between the reactant and product normalized to the number of Si atom deposited on the bulk and the cluster, is negative) then the reactivity of these screened precursors is tested again on the reconstructed and non-reconstructed Si(100) surfaces (Fig. 2). Among the studied 85 Si content precursors only two [$C_7H_{12}OSi$ -Methoxy-trivinyl-silane and $C_7H_9NSi$ - Benzyliminosilane] of them shows positive indications for ALD reactivity for Si deposition. On the other hand more than 10 precursors (ex: $C_{10}H_{22}N_2Si$, $SiCl_2H_2$) show positive indications for $SiO_2$ ALD reactivity). These predicted 10 precursors and the results will be reported in a forthcoming article. For a successive ALD deposition the chosen precursor should be reactive on the surface as well as it should be stable in the gas phase. From our vibrational analysis we have found that all these precursors are dynamically stable (no negative frequencies were observed). According to the literature, $C_7H_{12}OSi$ (13806759) is a stable precursor and its boiling point is 130°C (for more details see Ref. [17]). Similarly the existence of the $C_7H_9NSi$ (25933522) is also known experimentally and its stability is not reported in the literature (see ref. [18] and [19]).

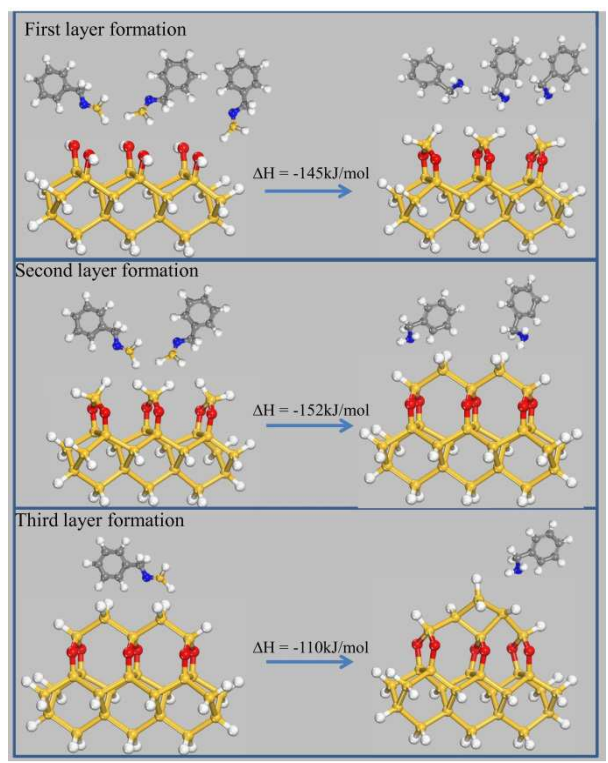

Fig1: Proposed schematic representation of reaction steps in three cycles of Si ALD using benzylimino-silane ($C_7H_9NSi$) on Si(100)-2$X$1 three-dimer surface.

**Table I**: List of Si content precursors used in the screening process. The Chemspider ID is given in the parenthesis. Precursors are grouped according to the molecular formula. All the precursors contain either H or C and H. In addiction one or more elements from Group V, Group VI and Group VII might be present.

| H | H,C | H,C,Group V | H,C,Group VI | H,C,Group VII |
|---|---|---|---|---|
| $H_4Si$ (22393), $H_6Si_2$ (66736), $H_8Si_3$ (122661). | $C_4H_{12}Si$ (6156), $C_5H_{10}Si$ (59499), $C_6H_{14}Si$ (63007), $C_6H_{18}Si_2$ (66675), $C_8H_{18}Si_2$ (76286), $C_8H_{20}Si$ (11919), $C_9H_{14}Si$ (63045), $C_{10}H_{10}Si$ (9227120), $C_{10}H_{16}Si$ (63058), $C_{10}H_{16}Si$ (263289), $C_{12}H_{12}Si$ (63083), $C_{14}H_{14}Si$ (123064), $C_{15}H_{15}Si$ (4887719), $C_{15}H_{15}Si$ (9404894), $C_{16}H_{16}Si$ (9520373), $C_{16}H_{16}Si$ (228965), $C_{17}H_{16}Si$ (502323), $C_{18}H_{16}Si$ (63117). | $C_4H_9NSi$ (74110), $C_5H_{14}SiN$ (14072949), $C_5H_{15}SiN$ (67521), $C_6H_{12}N_2Si$ (26905), $C_6H_{19}N_3Si$ (76493), $C_7H_9NSi$ (25933522), $C_7H_{19}NSi$ (63633), $C_8H_{22}N_2Si$ (10654667), $C_8H_{24}N_4Si$ (66803), $C_{10}H_{21}N_2Si$ (4909829) $C_{10}H_{22}N_2Si$ (9140753), $C_6H_{11}NSSi$ (511524), $C_9H_{23}NO_3Si$ (12933), $C_{13}H_{23}NOSi$ (313782). | $C_5H_{12}O_3Si$ (68503), $C_5H_{14}OSi$ (17017), $C_6H_{16}O_3SSi$ (19280), $C_6H_{18}OSi_2$ (23150), $C_7H_{12}OSi$ (13806759), $C_8H_{20}O_4Si$ (6270), $C_9H_{14}O_3Si$ (17131), $C_9H_{18}OSi$ (73227), $C_{12}H_{12}O_2Si$ (13100), $C_{14}H_{16}O_2Si$ (73339), $C_{18}H_{16}OSi$ (63119). $C_4H_9F_3O_3SSi$ (58839), $C_7H_{15}F_3O_3SSi$ (110158) | $ClH_3Si$ (55530), $Cl_2H_2Si$ (55266), $Cl_3HSi$ (23196), $Cl_4Si$ (23201), $Cl_6Si_2$ (75334), $FH_3Si$ (10328917), $F_2H_2Si$ (109934), $F_3HSi$ (122985), $F_4Si$ (22962), $F_6Si_2$ (123131), $BrH_3Si$ (55529), $Br_2H_2Si$ (123104), $Br_3HSi$ (74222), $Br_4Si$ (74225), $Br_6Si_2$ (13783143), $IH_3Si$ (10328917), $I_2H_2Si$ (122989), $I_3HSi$ (122986), $I_4Si$ (75335), $I_6Si_2$ (13783159), $CH_2Cl_4Si$ (14523), $CH_3Cl_3Si$ (6159), $C_2H_6Cl_2Si$ (6158), $C_3H_6Cl_2Si$ (29039), $C_3H_9ClSi$ (6157), $C_3H_9ISi$ (76879), $C_3H_9BrSi$ (68599), $C_4H_9F_3Si$ (480635), $C_4H_{11}ClSi$ (67897), $C_6H_{15}ClSi$ (13221), $C_6H_{15}ClSi$ (26908), $C_8H_{11}ClSi$ (12487), $C_{12}H_{10}Cl_2Si$ (6375), $C_{14}H_{14}ClSi$ (9308756), $C_{16}H_{15}F_2N_3Si$ (66326), $C_{18}H_{14}ClSi$ (4905153), $C_{18}H_{15}ClSi$ (6216). |

The adsorption of the precursor molecules on top of the hydroxyl group is addressed by optimizing the geometry of the total system (cluster plus precursor molecule). In all our calculation we have optimized a series of stable structures (from separated reactants to final products) along with a series of transition state connecting the various intermediate structures, to finally identify a candidate mechanism for the given chemical reaction. Finding a low-energy path from reactants to products allows us to state that the pathway is viable. However, we cannot ignore the possibility that a lower energy pathway may exist. According to our theoretical simulations we found the following two surface reactions during the first layer Si ALD using $C_7H_9NSi$ and $C_7H_{12}OSi$:

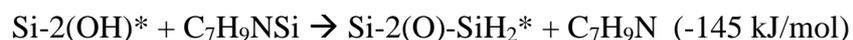

Si-2(OH)* + $C_7H_9NSi$ → Si-2(O)-SiH$_2$* + $C_7H_9N$  (-145 kJ/mol)

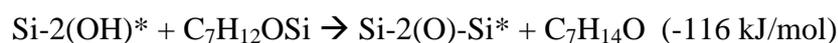

Si-2(OH)* + $C_7H_{12}OSi$ → Si-2(O)-Si* + $C_7H_{14}O$  (-116 kJ/mol)

For the second layer Si deposition:

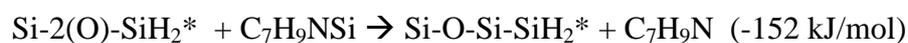

Si-2(O)-SiH$_2$* + $C_7H_9NSi$ → Si-O-Si-SiH$_2$* + $C_7H_9N$  (-152 kJ/mol)

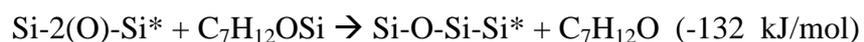

Si-2(O)-Si* + $C_7H_{12}OSi$ → Si-O-Si-Si* + $C_7H_{12}O$  (-132 kJ/mol)

For the third layer Si deposition:

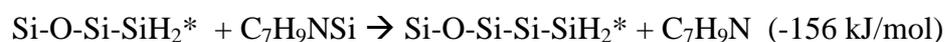

Si-O-Si-SiH$_2$* + $C_7H_9NSi$ → Si-O-Si-Si-SiH$_2$* + $C_7H_9N$  (-156 kJ/mol)

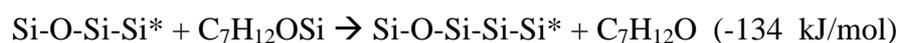

Si-O-Si-Si* + $C_7H_{12}OSi$ → Si-O-Si-Si-Si* + $C_7H_{12}O$  (-134 kJ/mol)

where the asterisks denote the surface moieties. The reaction energies are given in parenthesis and $C_7H_9N$ (benzylamine), $C_7H_{11}N$ (cyanocyclohexane) are the by-products of these gas-phase reactions. In both cases, the inter phase material becomes $SiO_2$ and the following second and third layers are Si. As mentioned in the beginning of the paper, it is challenging to remove the O from the interstitial layers, because of the strong covalent nature of the Si-O bond. The magnitude of the reaction energy suggests that these reactions are thermodynamically feasible and the reported reactions are favourable path for the Si-ALD. In an ALD method based on a nethoxy-trivinyl-silane precursor, the breaking of a Si-OR (organic species) bond by a hydroxyl group occurs via a simultaneous two steps reaction: 1) the chemisorption of the silane onto a hydroxyl terminated surface, and 2) the subsequent rearrangement of H with surface-bound Si-O/Si-Si. The presence of a basic amino group on a nethoxy-trivinyl-silane might allow reactions (1) and (2) to happen without added catalyst.

The reactivity of these screened precursors are again tested on the reconstructed (fig 2a), and non-reconstructed (see fig 2a) hydroxylated Si(100) surface and we found that the reaction energies predicted by the cluster model are closer to the surface models. The calculated reaction energies for the first, second, and third layers formation using $C_7H_9NSi$ precursor on the reconstructed/non-reconstructed hydroxylated Si(100) surface are -125/-114, -135/-129, -133/-122 kJ/mol, respectively. When we used $C_7H_{12}OSi$ as a precursor the calculated reactions energies are -102/-100 and -128/-112, -132/-114 kJ/mol; for first, second and third layers formation, respectively. It should be noticed that due to the similar type of reactions in the second and third layers formation, the formation energy does not change considerably. The magnitude of the formation energy on both the cluster and the surface model suggested that $C_7H_9NSi$ precursor is the best choice for the Si deposition. In order to substantiate our prediction, experimental verification is needed. Our experimental team is working on the synthesis of the proposed precursors in order to make the Si-ALD deposition.

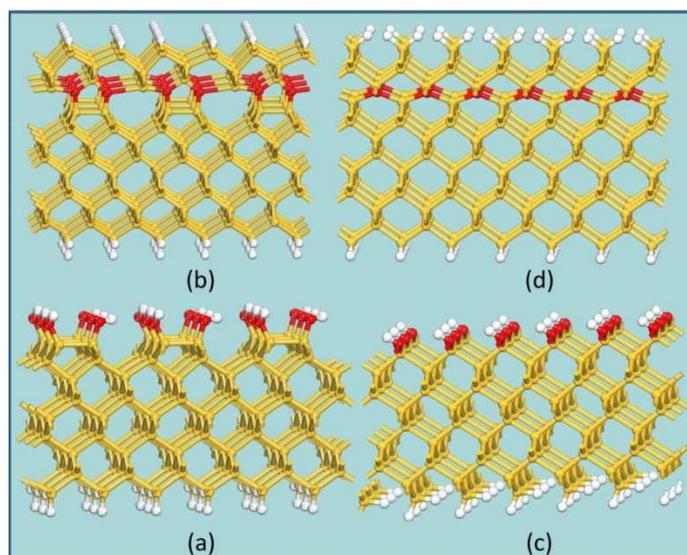

Fig 2: Optimized surface models for absorption of three layers of Si on hydroxylated Si(100) surface. Reconstructed (a), non-reconstructed (c) OH terminated Si(100) surface and the corresponding three layers of Si deposited (b and d) surface models. The yellow, red and white atoms denote silicon, oxygen, and hydrogen atoms, respectively.

**Conclusion**

The initial surface reaction mechanism of ALD-grown Si on the hydroxylated Si(100)-2 X 1 surface using several potential silicon precursors were screened using state-of-the-art density functional calculations. According to our theoretical simulations we proposed that it is possible to deposit Si on OH terminated reconstructed/non-reconstructed Si(100) surface and the formation energies suggest that $SiO_2$ deposition is much easier than Si deposition at the initial stage of ALD. Similarly the calculations on both cluster and surface models clearly indicate that the interface between the substrate and the deposited Si films forms a Si-O layer.

**Acknowledgments**

The authors gratefully acknowledge the Research Council of Norway (Grant No:143408) for financial support. PV acknowledges the Research Council of Norway for providing the computer time at the Norwegian supercomputer facilities (under the project number NN2875k).